\documentclass{aa}  
\usepackage[varg]{txfonts}
\usepackage{graphicx}
\usepackage{amssymb,amsmath}
\usepackage{booktabs,multirow}
\usepackage{bm}
\usepackage{natbib}
\bibpunct{(}{)}{;}{a}{}{,}

\usepackage{color}

\begin{document}

\title{Relative field line helicity of a large eruptive solar active region}

\author{K. Moraitis$^1$ \and S. Patsourakos$^1$ \and A. Nindos$^1$}
\institute{Physics Department, University of Ioannina, Ioannina GR-45110, Greece}

\date{Received ... / Accepted ...}

\abstract{Magnetic helicity is a physical quantity of great importance in the study of astrophysical and natural plasmas. Although a density for helicity cannot be defined, a good proxy for it is field line helicity. The appropriate quantity for use in solar conditions is relative field line helicity (RFLH).}{This work aims to study in detail the behaviour of RFLH, for the first time, in a solar active region (AR).}{The target active region is the large, eruptive AR 11158. In order to compute RFLH and all other quantities of interest we use a non-linear force-free reconstruction of the AR coronal magnetic field of excelent quality.}{We find that the photospheric morphology of RFLH is quite different than that of the magnetic field or of the electrical current, and this is not sensitive to the chosen gauge in the computation of RFLH. The value of helicity experiences a large decrease, $\sim 25\%$ of its pre-flare value, during an X-class flare of the AR, a change that is also depicted in the photospheric morphology of RFLH. Moreover, the area of this change coincides with the area that encompasses the flux rope, the magnetic structure that later erupted.}{The use of RFLH can provide important information about the value and location of the magnetic helicity expelled from the solar atmosphere during eruptive events.}


\titlerunning{Relative field line helicity of AR 11158}
\authorrunning{Moraitis et al.}

\maketitle

\section{Introduction}
\label{sect:introduction}

Magnetic helicity is a quantity that represents an important restriction in the evolution of astrophysical plasmas in the magnetohydrodynamics (MHD) description, since it is conserved in ideal MHD \citep{woltjer58}, and approximately conserved in non-ideal conditions \citep{taylor74,pariat15}. Magnetic helicity indicates the complexity of a magnetic field, as it measures the twist, writhe, and interlinking of the magnetic field lines. Mathematically, it is defined as the volume integral $H=\int_V \mathbf{A}\cdot \mathbf{B}\,{\rm d}V$, where $\mathbf{B}$ is the magnetic field in the volume $V$, and $\mathbf{A}$ the corresponding vector potential such that $\mathbf{B}=\nabla\times \mathbf{A}$. Helicity thus depends on the chosen gauge for $\mathbf{A}$ unless the volume of interest is bounded by a magnetic flux surface. This condition is not met in the Sun where magnetic flux is exchanged between the photosphere and the solar interior. The appropriate quantity in this case is relative magnetic helicity \citep{BergerF84}, which is defined with the help of a reference magnetic field.

An important application of the conservation property of helicity in the Sun
is to the initiation of coronal mass ejections (CMEs). These eruptions usually involve the expulsion of highly-twisted, i.e., helical, magnetic structures, the flux ropes \citep[FRs,][]{green18,patsourakos20}. According to a popular scenario \citep{rust94,low94}, CMEs are the means for the corona to remove the constantly accumulating helicity from the solar interior. In cases like this it would be desirable to be able to pinpoint the locations on the Sun, or in the solar corona, where helicity is more important.

The classical and relative helicities however are non-local quantities, due to their dependence on the vector potential, which means that the respective densities lack a proper definition. The most meaningful way to define a helicity density is with field line helicity \citep[FLH,][]{antiochos87,berger88}. Field line helicity is given by the line integral $h=\int_C\,\mathbf{A} \cdot {\rm d}\bm{l}$ along a field line $C$. FLH is a function of the field lines, which expresses the flux of the magnetic field through the surface bounded by the field line, when this is closed, otherwise, it has no direct physical meaning.

The appropriate generalization of FLH for use at the solar environment, the relative field line helicity (RFLH), was only recently developed \citep{yeates18,moraitis19}. These works defined two distinct forms for relative FLH which differ in the used gauge. Both forms however, accurately reproduce relative helicity when all field lines that pierce the boundary of the volume are considered.

Field line helicity, in its classical or relative form, has been employed in various solar applications. In one application, FLH was used to trace the connectivity changes caused by reconnection in a magnetic field setup which simulated coronal loops \citep{russel15}. In another example, \citet{yeates16} have applied FLH to the global magnetic field of the Sun in order to study the distribution of helicity in the solar corona. An interesting application of FLH was in the determination of the FRs location in an approximation of the solar corona, through a threshold in FLH \citep{lowder17}. The properties of FLH have also been examined theoretically in \citet{aly18}. Most of the applications of FLH and RFLH so far are therefore either in idealized situations, or to simulations of the actual solar environment.

A first, more realistic application of RFLH to observed solar ARs was made in \citet{moraitis19}. Using a non-linear force-free (NLFF) reconstruction of the coronal magnetic field of a solar active region at a specific instant, the authors examined the relation of RFLH with the helicity flux density \citep[e.g.,][]{pariat05}. They showed that the two quantities exhibit overall similar photospheric spatial distributions, while some minor differences were also observed. This work highlighted the potential of using RFLH in actual solar conditions.

In this work, we go one step further, as our purpose is to study the behaviour of relative field line helicity in a large observed active region (AR) over an extended time interval. In Sect.~\ref{sect:method} we define all quantities of interest, and we describe the observational data, as well as the implementation details, that we use. In Sect.~\ref{sect:results} we present the results of the application of RFLH in the chosen AR, focusing on a specific large flare. Finally, in Sect.~\ref{sect:discussion} we summarize and discuss the results of the paper.

\section{Methodology}
\label{sect:method}

\subsection{Active region selection}
\label{sect:data}

The active region chosen for the first detailed study of the behaviour of RFLH in solar conditions is AR 11158. The reasons for this choice are many. First, there is an extensive literature about this AR, with many different of its characteristics examined \citep[e.g.,][]{sun12,tgl13}. Second, during the passage of the AR from the solar disk it exhibited a range of behaviours, with most notable, the numerous flares (up to the X-class) and the CMEs that accompanied many of them. Third, a high-quality reconstruction of the AR's coronal magnetic field, that is necessary in our computations, was available to us \citep{thalmann19}.

We employed two data products from the Solar Dynamic Observatory \citep[SDO,][]{pes12} in this work. These were the \texttt{sharp$\_$cea$\_$720s} data series from the Helioseismic and Magnetic Imager \citep[HMI,][]{sche12} instrument and the \texttt{lev1$\_$uv$\_$24s} one at 1600~\r{A} from the Atmospheric Imaging Assembly (AIA) instrument \citep{aiapaper}. The time interval that we examined was between 12 February 2011, 00:00 UT and 16 February 2011, 00:00 UT. In this interval AR 11158 exhibited two M-class flares, an M6.6 on 13 February with peak time at 17:38 UT and an M2.2 on 14 February at 17:26 UT, and the X2.2 flare on 15 February at 01:56 UT, all of which were eruptive.

\subsection{Quantities of interest}
\label{sect:rmflh1}

\subsubsection{Relative magnetic helicity}

The appropriate magnetic helicity for the Sun is relative magnetic helicity \citep{BergerF84}. For a three-dimensional (3D) magnetic field $\mathbf{B}$ in a finite volume $V$, this is given by the \citet{fa85} formula
\begin{equation}
H_r=\int_V (\mathbf{A}+\mathbf{A}_\mathrm{p})\cdot (\mathbf{B}-\mathbf{B}_\mathrm{p})\,{\rm d}V.
\label{helr}
\end{equation}
The magnetic field $\mathbf{B}_\mathrm{p}$ serves as a reference field, and usually it is chosen to be potential, while $\mathbf{A}$, $\mathbf{A}_\mathrm{p}$ are the respective vector potentials of the two magnetic fields. Relative magnetic helicity is independent of the chosen gauges of the two vector potentials, as long as the normal components of the magnetic fields are the same on the boundary of the volume, $\partial V$. This condition can be written as
\begin{equation}
\left. \hat{n}\cdot \mathbf{B} \right|_{\partial V}=\left. \hat{n}\cdot \mathbf{B}_\mathrm{p} \right|_{\partial V},
\label{helc}
\end{equation}
with $\hat{n}$ denoting the outward-pointing unit normal on $\partial V$.

\subsubsection{Energies}

Although we are not interested in the evolution of the various energies in this work, there are energy-related parameters that help us identify the quality of a given magnetic field. We introduce these parameters in the following.

The energy of a magnetic field $\mathbf{B}$ is given by the well-known relation
\begin{equation}
E=\frac{1}{2\mu_0}\int_V \mathbf{B}^2 {\rm d}V,
\label{eq:nrg}
\end{equation}
where $\mu_0$ is the magnetic permeability of the vacuum. A useful decomposition of this energy is based on the unique splitting of the magnetic field into potential and current-carrying components, $\mathbf{B}=\mathbf{B}_\mathrm{p}+\mathbf{B}_\mathrm{j}$. Following \citet{val13}, each component can be further divided into solenoidal and non-solenoidal parts, as $\mathbf{B}_\mathrm{p}=\mathbf{B}_\mathrm{p,s}+\mathbf{B}_\mathrm{p,ns}$ and $\mathbf{B}_\mathrm{j}=\mathbf{B}_\mathrm{j,s}+\mathbf{B}_\mathrm{j,ns}$. By defining the energy of each component from Eq.~(\ref{eq:nrg}) we end up with the following decomposition
\begin{equation}
E=E_\mathrm{p,s}+E_\mathrm{p,ns}+E_\mathrm{j,s}+E_\mathrm{j,ns}+E_\mathrm{mix},
\label{eqdecomp}
\end{equation}
where $E_\mathrm{mix}$ denotes the energy corresponding to all the cross terms. For a perfectly solenoidal magnetic field, Eq.~(\ref{eqdecomp}) reduces to the more familiar form $E=E_\mathrm{p,s}+E_\mathrm{j,s}$, where $E_\mathrm{j,s}$ can be identified as the free energy.

The non-solenoidal parts of the magnetic fields and the respective terms in Eq.~(\ref{eqdecomp}) are non-physical and stem solely from numerical reasons. Collecting these terms together, leads to the definition of the energy
\begin{equation}
E_\mathrm{div}=E_\mathrm{p,ns}+E_\mathrm{j,ns}+|E_\mathrm{mix}|,
\label{eqediv}
\end{equation}
which is a measure of how much a numerical model of a magnetic field deviates from solenoidality; lower values of $E_\mathrm{div}$, and consequently of the energy ratio $E_\mathrm{div}/E$, indicate a less solenoidal magnetic field. Regarding helicity computations, \citet{valori16} have set a threshold of $E_\mathrm{div}/E\lesssim 0.08$ for helicity values to be reliable, which was later refined to $E_\mathrm{div}/E\lesssim 0.05$ by \citet{thalmann19a},  where the various energies for the same AR 11158 were discussed.

Another energy ratio that is related to the quality of a magnetic field with respect to its solenoidality is $|E_{\rm mix}|/E_{\rm j,s}$, which shows the importance of the cross term in Eq.~(\ref{eqdecomp}) relative to the free energy. Values of $|E_{\rm mix}|/E_{\rm j,s}\lesssim 0.35$ are related with magnetic fields of higher quality \citep{thalmann20}.

\subsubsection{Relative field line helicity}

\subsubsection*{a. Unconstrained gauge}

The most general expression for the field line helicity that corresponds to relative magnetic helicity was derived in \citet{moraitis19}, and it is reviewed here for completeness. Relative field line helicity (RFLH), $h_r$, can be considered as the density of relative magnetic helicity per unit magnetic flux. This follows from the relation
\begin{equation}
H_r=\oint_{\partial V} h_r\,{\rm d}\Phi,
\label{flhhel}
\end{equation}
where ${\rm d}\Phi=\left| \hat{n}\cdot\mathbf{B} \right|\,{\rm d}S$ is the elementary magnetic flux on the boundary (${\rm d}S$: elementary area). Equation~(\ref{flhhel}) results directly from Eq.~(\ref{helr}) after expanding the integrand of the latter and splitting the volume element along the field lines of $\mathbf{B}$ as ${\rm d}V={\rm d}\mathbf{S}\cdot {\rm d}\bm{l}$, or, of $\mathbf{B}_\mathrm{p}$ as ${\rm d}V={\rm d}\mathbf{S}\cdot {\rm d}\bm{l}_\mathrm{p}$ (${\rm d}\bm{l}$, ${\rm d}\bm{l}_\mathrm{p}$: elementary lengths along respective field lines). The only assumption made in deriving Eq.~(\ref{flhhel}) is the reasonable smoothness of the magnetic field, and especially the lack of null points where field lines are discontinuous. Additionally, this relation is useful when most of the field lines of $\mathbf{B}$ are connected to the boundary at both ends, i.e., when the number of closed field lines inside the volume is limited.

\begin{figure}[h]
\centering
\includegraphics[width=0.42\textwidth]{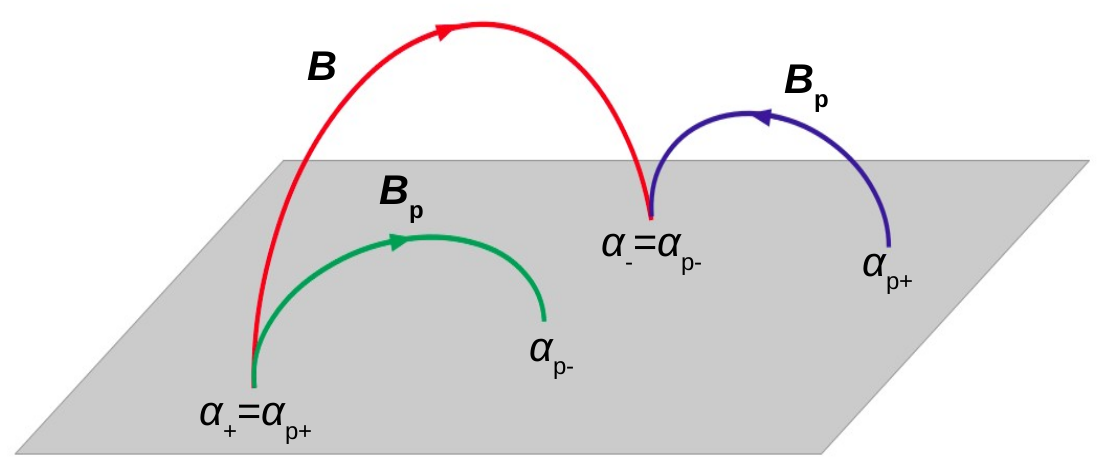}
\caption{Sketch of the field lines involved in the definition of relative field line helicity.}
\label{flhfig}
\end{figure}

As its name indicates, RFLH is a function of the field lines that assigns a single value to each field line (pair of footpoints), and so, it can be viewed as a 2D map that depends on the location on the boundary. Denoting the part of the boundary where magnetic flux enters into (leaves) the volume as $\partial V^+$ ($\partial V^-$), and the footpoints of a generic field line of $\mathbf{B}$ as $\alpha_+\in \partial V^+$, $\alpha_-\in \partial V^-$ (and as $\alpha_{p+}\in \partial V^+$, $\alpha_{p-}\in \partial V^-$ for $\mathbf{B}_\mathrm{p}$, respectively), then RFLH is given by any of the following expressions
\begin{equation}
h_r^+=\int_{\alpha_+}^{\alpha_-}\,(\mathbf{A}+\mathbf{A}_\mathrm{p}) \cdot {\rm d}\bm{l} - \int_{\alpha_+}^{\alpha_{p-}}\,(\mathbf{A}+\mathbf{A}_\mathrm{p}) \cdot {\rm d}\bm{l}_\mathrm{p},
\label{flhdef}
\end{equation}
or
\begin{equation}
h_r^-=\int_{\alpha_+}^{\alpha_-}\,(\mathbf{A}+\mathbf{A}_\mathrm{p}) \cdot {\rm d}\bm{l} - \int_{\alpha_{p+}}^{\alpha_{-}}\,(\mathbf{A}+\mathbf{A}_\mathrm{p}) \cdot {\rm d}\bm{l}_\mathrm{p},
\label{flhdefm}
\end{equation}
or, by their average, as
\begin{equation}
h_r^0=\frac{1}{2}\left( h_r^+ + h_r^-\right).
\label{flhdef0}
\end{equation}
The first expression involves the integration along field lines of $\mathbf{B}$ and $\mathbf{B}_\mathrm{p}$ that have common their positive-polarity footpoints, $\alpha_{+}=\alpha_{p+}$, as shown by the red and green field lines in Fig.~\ref{flhfig}. When using $h_r^+$ in Eq.~(\ref{flhhel}), only the positive-polarity part of the boundary, $\partial V^+$, should be considered, so that each field line is counted once.

Similarly, if the field lines of the two fields have in common the negative-polarity footpoints, $\alpha_{-}=\alpha_{p-}$, then the RFLH expression given by Eq.~(\ref{flhdefm}) should be used along the $\partial V^-$ boundary (red and blue field lines in Fig.~\ref{flhfig}). When the whole boundary is considered, the respective RFLH, Eq.~(\ref{flhdef0}), involves all field lines shown in Fig.~\ref{flhfig}. In all cases the RFLH is expressed as a line integral along the field lines of $\mathbf{B}$ relative to the same quantity along those of $\mathbf{B}_\mathrm{p}$, thus justifying its name.\\

\subsubsection*{b. Berger \& Field gauge}

The expressions for RFLH given above make no assumption on the gauge of the vector potentials. RFLH however is a gauge-dependent quantity, as it can easily be seen from any of Eqs.~(\ref{flhdef})-(\ref{flhdef0}), and this can be exploited in order to derive simpler expressions for the RFLH. A commonly-used gauge for the two vector potentials is given by the relation
\begin{equation}
\left. \hat{n}\times\mathbf{A}_\mathrm{p} \right|_{\partial V} = \left. \hat{n}\times\mathbf{A} \right|_{\partial V}.
\label{eq:gaugeb2}
\end{equation}
This gauge condition can be fulfilled with the tangential components of $\mathbf{A}_\mathrm{p}$ determined from those of $\mathbf{A}$, while the latter can be in any gauge. The condition given by Eq.~(\ref{eq:gaugeb2}) leads to the elimination of the cross terms in the integrand of Eq.~(\ref{helr}), which then reads 
\begin{equation}
H_r=\int_V \mathbf{A}\cdot \mathbf{B}\,{\rm d}V-\int_V \mathbf{A}_\mathrm{p}\cdot \mathbf{B}_\mathrm{p}\,{\rm d}V.
\label{bfgauge}
\end{equation}
This gauge was used in the original definition of relative magnetic helicity by \citet{BergerF84}. Starting from Eq.~(\ref{bfgauge}), \citet{yeates18} defined the following relative field line helicity
\begin{equation}
\tilde{h}_r^{+}= \int_{\alpha_+}^{\alpha_-}\,\mathbf{A} \cdot {\rm d}\bm{l} - \int_{\alpha_+}^{\alpha_{p-}}\,\mathbf{A}_\mathrm{p} \cdot {\rm d}\bm{l}_\mathrm{p},
\label{flhyp}
\end{equation}
written here for the positive-polarity boundary, with analogous to Eqs.~(\ref{flhdefm}) and (\ref{flhdef0}) relations holding for the other expressions, $\tilde{h}_r^-$, $\tilde{h}_r^0$. These relative RFLHs are constructed similarly to the RFLHs in the uncontsrained gauge, they employ however a specific gauge condition.


\subsection{Implementation}

\subsubsection{Magnetic field modelling}

All physical quantities that interest this work require as input the 3D magnetic field in the coronal volume. The 3D field at each instant is reconstructed from the corresponding observed HMI magnetogram with a NLFF method. This uses a weighted optimization approach \citep{wieg10}, after pre-processing the horizontal components of the field on the photosphere so that it becomes more compatible with the force-free assumption, and additionally, smoothing of the original vector magnetogram. The full details of the method can be found in \citet{thalmann19} where this reconstruction was first used. 

We only mention here that the original HMI data were slightly cropped so that to leave quiet-Sun regions outside, and then rebinned by a factor of four, to the resolution of $2\arcsec$ per pixel. The resulting NLFF field occupies the volume $215\,{\rm Mm}\times 130\,{\rm Mm}\times 185\,{\rm Mm}$, and it is discretized by $148\times 92 \times 128$ grid points. In total, there are 115 snapshots with the typical cadence of 1~hr, except around the M6.6 and the X2.2 flares, when HMI's highest cadence of 12 minutes was used.

\begin{figure}[h]
\centering
\includegraphics[width=0.48\textwidth]{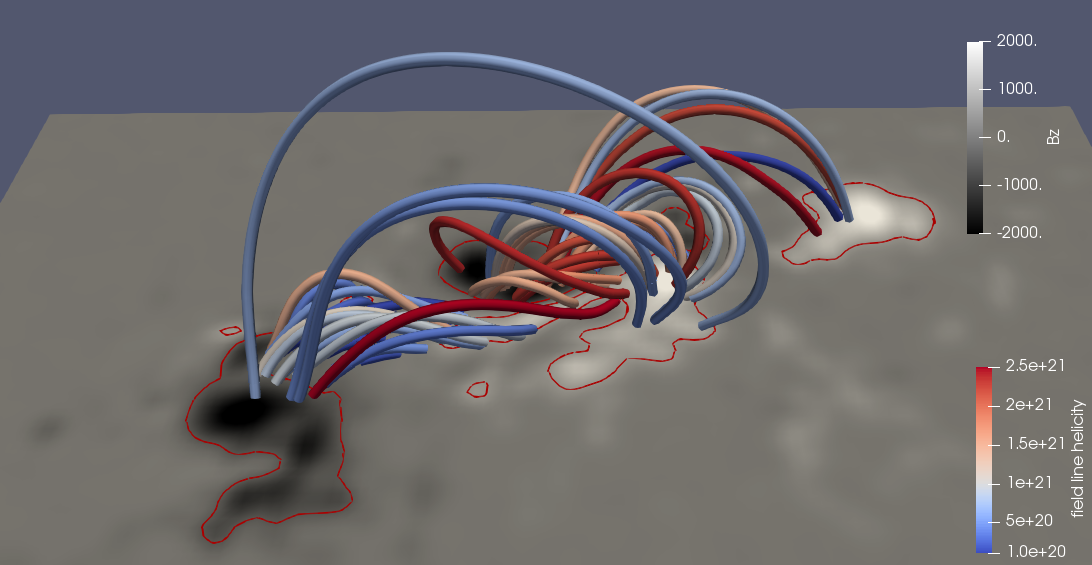}
\caption{Morphology of the 3D magnetic field of AR 11158 on 01:11 UT of 15 February 2011, for the NLFF extrapolation used in this work. Field lines are coloured according to the value of RFLH, which is positive for all selected field lines, and they are overplotted on the photospheric distribution of the vertical magnetic field, with the red contours corresponding to the values $|B_z|=500\,{\rm G}$.}
\label{datafig}
\end{figure}

The morphology of the extrapolated magnetic field for a snapshot $\sim$~1~hr before the X-class flare is shown in Fig.~\ref{datafig}. A low-lying flux rope surrounded by an arcade field can be identified there, as is already known by other works \citep[e.g.,][]{sun12,nindos12}. Moreover, Fig.~\ref{datafig} shows that the RFLH values of the flux rope are higher than in the arcade field.

One of the reasons for choosing the specific AR for this study is the excellent quality of the reconstruction of the AR's 3D magnetic field. The quality of the field is intended as the accuracy in fulfilling the assumptions of the NLFFF method, as quantified by the low values of divergence-, and force-freeness reported in \citet{thalmann19}. The respective parameters used by the authors are the average absolute fractional flux increase and the average current-weighted angle between the current and the magnetic field \citep{wheatland00}.

Besides these parameters the high level of divergence-freeness of the magnetic field is further shown with the following three parameters. The first is the recently-proposed modification of the fractional flux increase which, unlike the original parameter, is independent of the mesh size \citep{gilchrist20}. This is quite low, with a mean value over all the 115 snapshots of $\langle|f_d|\rangle=(1.50\pm 0.06)\times 10^{-9}\, \mathrm{m}^{-1}$. Additionally, the energy ratio proposed by \citet{valori16} is well below the threshold of 0.08, or even of 0.05, with a value of $E_{\rm div}/E=(6.1\pm 0.3)\times 10^{-3}$. The other energy ratio \citep{thalmann20} is also much lower than the respective limit of 0.35, with a value of $|E_{\rm mix}|/E_{\rm j,s}=0.051\pm 0.006$. Such high quality of the reconstructed magnetic field thus quarantees the reliability of the subsequent helicity computations.

\subsubsection{Numerical computations}

The computation of all the physical quantities can be performed in a three-step process, once the magnetic field is provided by the NLFF method. The first step is the computation of the potential field that satisfies the condition of Eq.~(\ref{helc}). This is derived by numerically solving Laplace's equation. The second step involves the computation of the vector potentials from the respective magnetic fields. This is done by adopting the \citet{devore00} gauge as this was modified by \citet{val12}. The full details involved in these two steps are described in \citet{moraitis14}.

The third step is required only for the RFLH computations, and it involves two different field line integrations, one for $\mathbf{B}$, and one for $\mathbf{B}_\mathrm{p}$. These integrations are performed with the fast and robust method that is described in \citet{moraitis19}. A difference with that work is that here we consider only the field lines with (at least one) footpoint on the photospheric part of the boundary. We thus neglect the lateral and top parts of the boundary, as our experience has shown that a very small piece of information is lost. Once the field lines are obtained, we compute the RFLH expression in the unconstrained gauge, $h_r^0$, from Eq.~(\ref{flhdef0}).

For the computation of the RFLH in the Berger \& Field gauge, $\tilde{h}_r^0$, the three-step process is slightly different than the one just described. The reason is that the computation is performed with the code provided by \citet{yeates18}, which uses a different calculation for the potential magnetic field, and treats all vector quantities in a staggered grid instead of being colocated.

\section{Results}
\label{sect:results}

\subsection{RFLH morphology}
\label{sect:res1}

\begin{figure*}[h]
\centering
\includegraphics[width=0.32\textwidth]{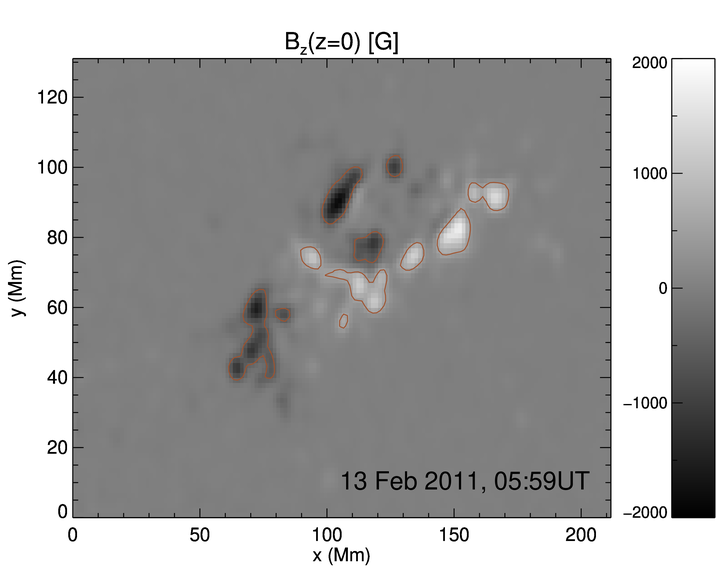}%
\includegraphics[width=0.32\textwidth]{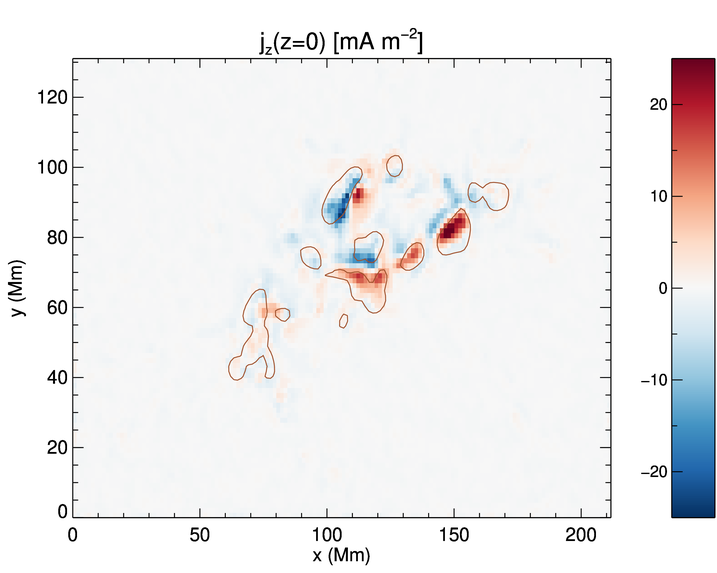}%
\includegraphics[width=0.32\textwidth]{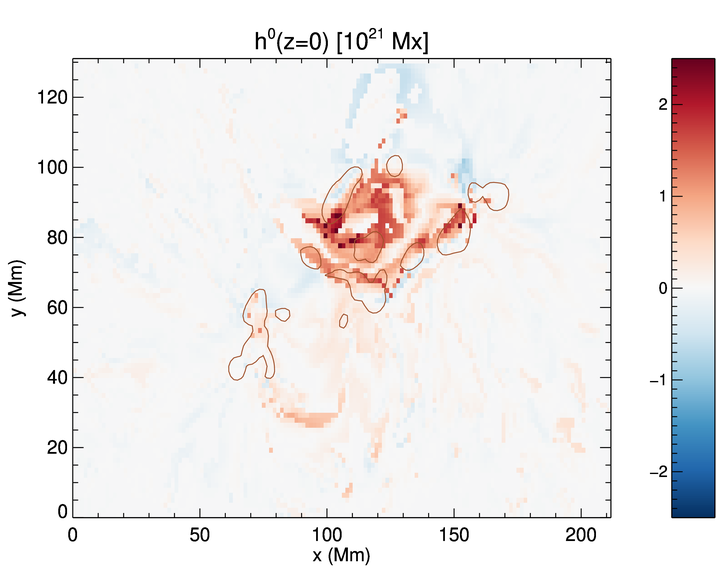}\\
\includegraphics[width=0.32\textwidth]{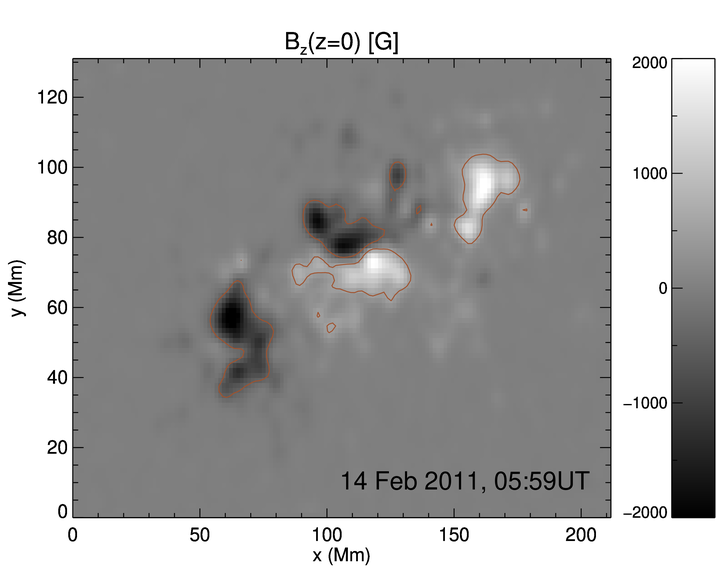}%
\includegraphics[width=0.32\textwidth]{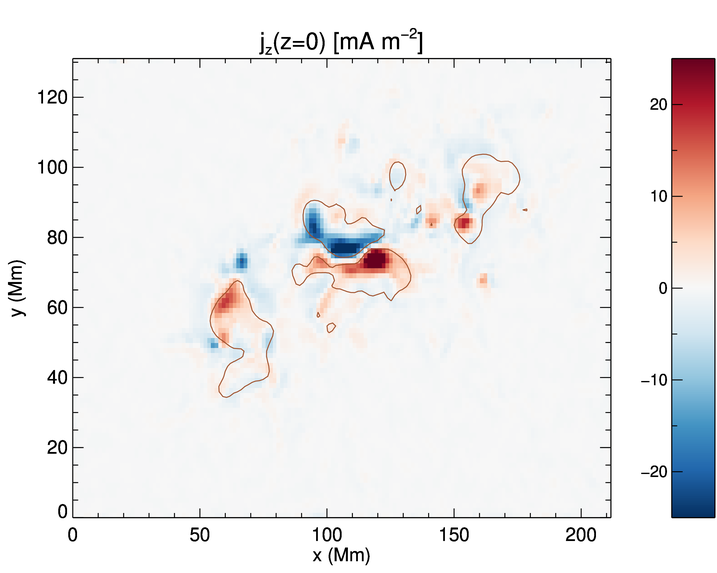}%
\includegraphics[width=0.32\textwidth]{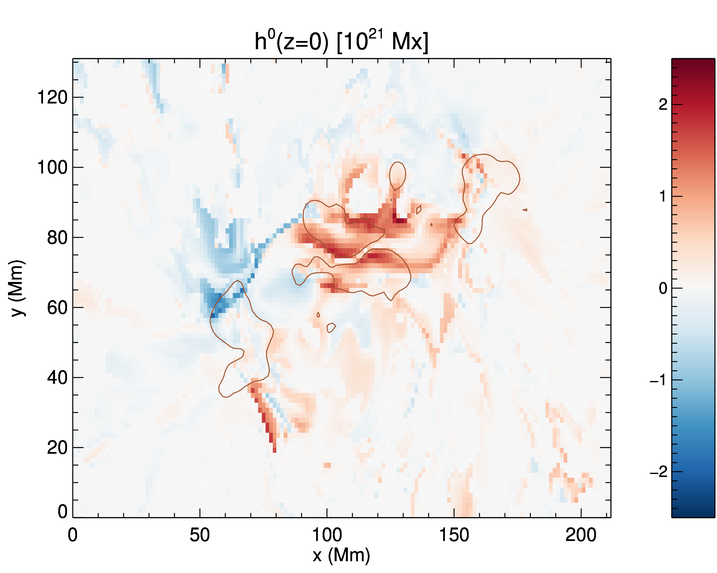}\\
\includegraphics[width=0.32\textwidth]{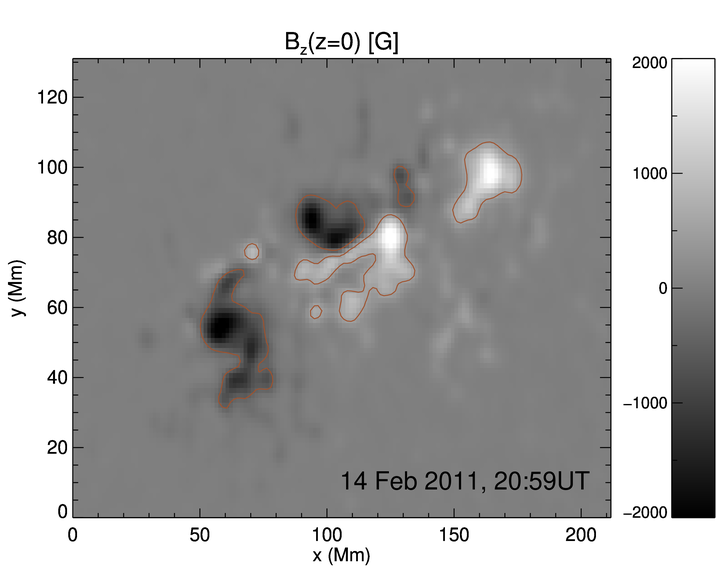}%
\includegraphics[width=0.32\textwidth]{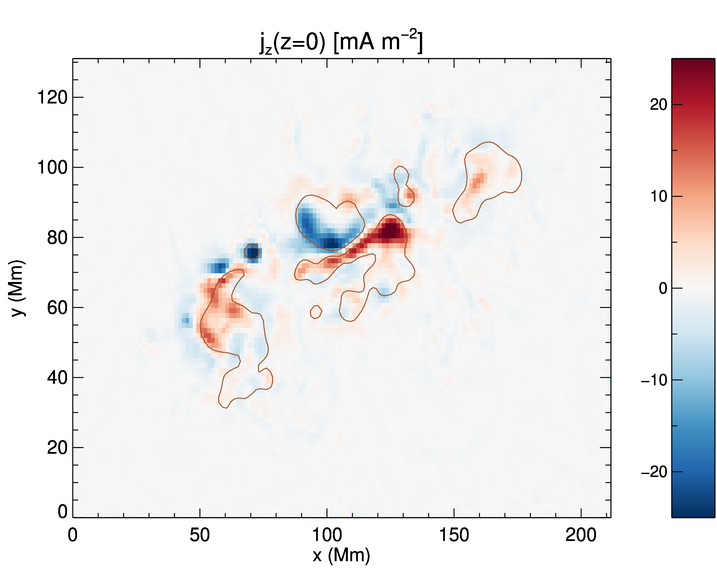}%
\includegraphics[width=0.32\textwidth]{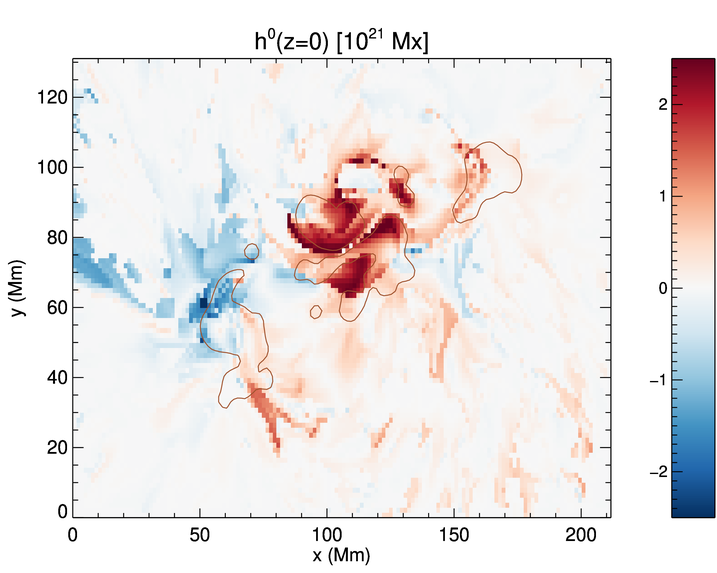}\\
\includegraphics[width=0.32\textwidth]{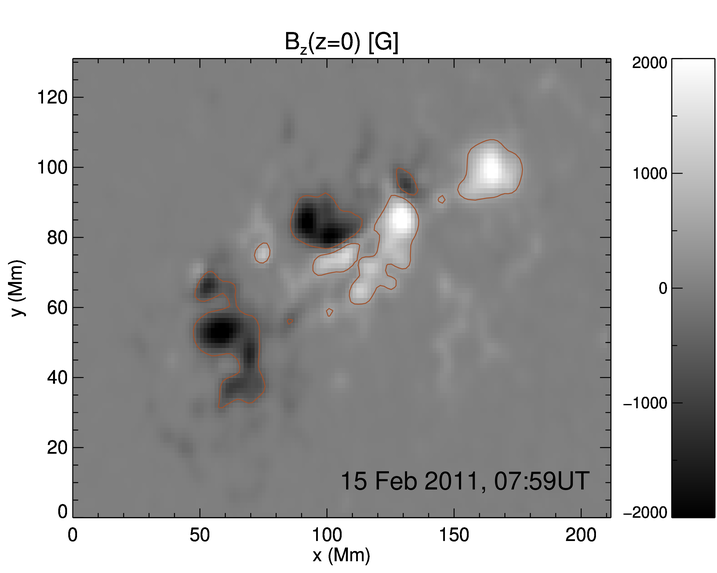}%
\includegraphics[width=0.32\textwidth]{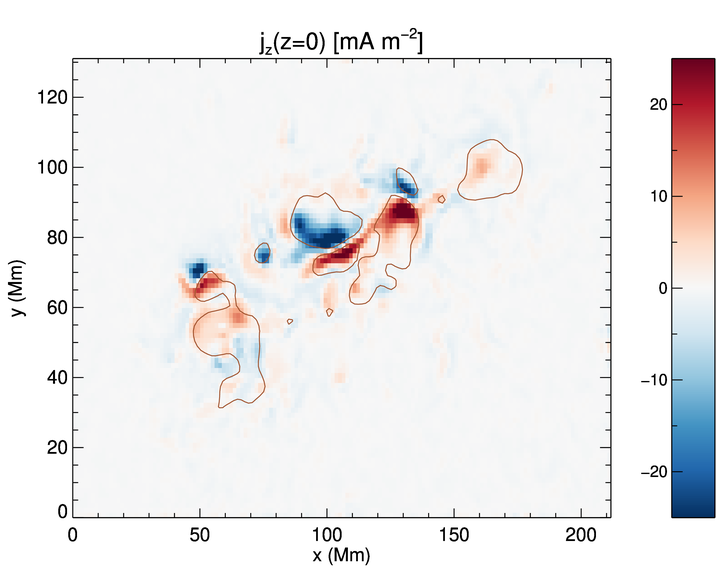}%
\includegraphics[width=0.32\textwidth]{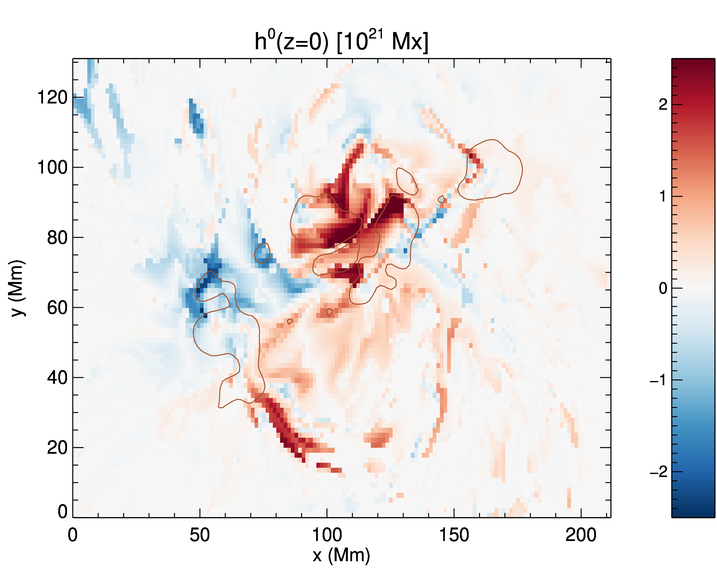}%
\caption{Photospheric maps of vertical magnetic field, $B_z$ (left), vertical electrical current, $j_z$ (middle), and RFLH (right) in AR 11158 for four snapshots during the studied interval. The red contours correspond to the vertical magnetic field values $|B_z|=500\,{\rm G}$ at each instant.}
\label{flhfig1}
\end{figure*}

Following the methodology in Sect.~\ref{sect:method} we calculate all quantities of interest for all available snapshots of the NLFF magnetic field model. In Fig.~\ref{flhfig1} we show four snapshots from the evolution of the morphology of RFLH on the photospheric plane, $z=0$, in the unconstrained gauge, $h_r^0$, and also, of the vertical magnetic field, $B_z$, and of the vertical electrical current, $j_z=\left( \nabla\times\mathbf{B} \right)_z/\mu_0$, also on the photospheric level. Each snapshot provides an example of the field configuration in each of the four intervals in which the major flares divide our study interval into: one before the M6.6 flare, one between the two M-class flares, one between the M2.2 and the X2.2 flares, and one after the X2.2 flare.

The morphology of the vertical magnetic field evolved from two initial nearby bipoles that converged into a highly sheared bipolar flux distribution internal to the outer polarities \citep[e.g.,][]{chintzoglou13}. The vertical current distribution primarily develops around the polarity inversion lines (PILs), and it follows to a large degree the evolution of the magnetic field, as the comparison with the respective contours shows in Fig.~\ref{flhfig1}. The RFLH morphology is however different than the other distributions. It is more extended, and it has a much less coincidence with the magnetic field contours. The sign of the RFLH is mostly positive, in agreement with the AR helicity sign. The highest (positive) RFLH values are located around the core polarities of the AR. This central part of RFLH becomes more intense until around the time of the X-class flare, and finally relaxes to lower values. The behaviour of RFLH around the flare is examined in more detail in Sect.~\ref{sect:res3}. The lowest RFLH values are found to the east periphery of the AR (left in the respective panels of Fig.~\ref{flhfig1}), with only a small coincidence with the nearby magnetic field polarity.

\begin{figure}[h]
\centering
\includegraphics[width=0.48\textwidth]{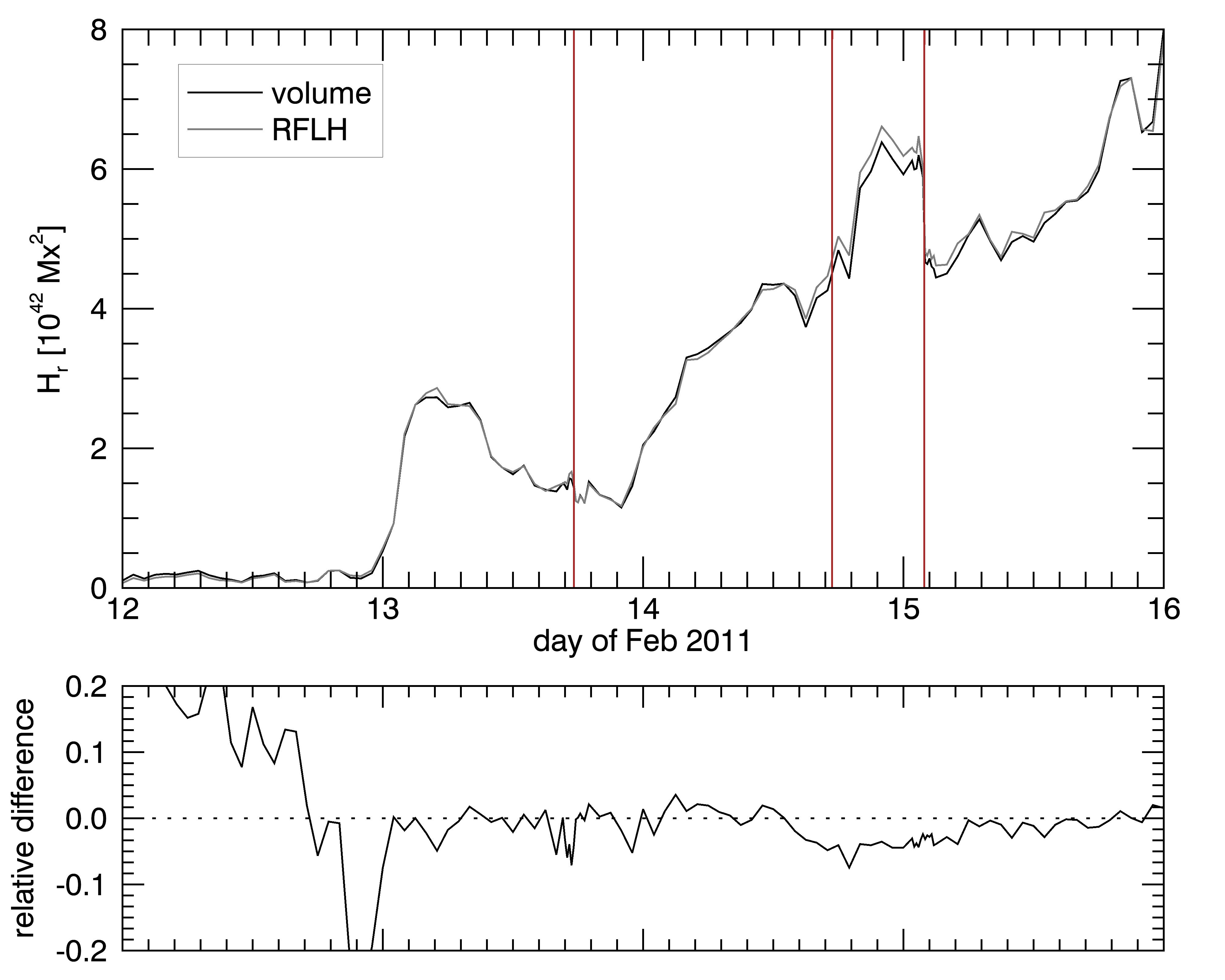}
\caption{Evolution of relative helicity in AR 11158 as calculated by the volume (Eq.~(\ref{helr}), black line), and the RFLH methods (Eq.~(\ref{flhhel}), grey line) in the top panel, and of their relative difference in the bottom, during the whole studied interval. The vertical lines denote the times of the two M-, and of the X-class flares.}
\label{helplots}
\end{figure}

The corresponding evolution of the relative helicity budgets, as calculated by the volume method of Eq.~(\ref{helr}), and of the RFLH method of Eq.~(\ref{flhhel}) on the full magnetogram, is shown in Fig.~\ref{helplots} with the black and grey curves, respectively. We note that from 13 February onwards, the helicities obtained with the two methods agree to a large extent. We quantify this agreement in the bottom panel of Fig.~\ref{helplots} where we notice that the relative difference between the two curves is within $\lesssim 5\%$, with an average absolute value of 2.2\%. In calculating these numbers we omit the day of February 12 when the values of helicity are very low and consequently the relative difference much higher than in the following days. The closeness of the two helicity curves indicates that the RFLH method works as expected, since it exhibits similar levels of agreement as in \citet{moraitis19}. It also shows that most of the field lines contributing to helicity are indeed rooted in the phothosphere rather than on the lateral boundaries. Both methods indicate a large drop in helicity during the X-class flare, which we further discuss in the following sections.

\begin{figure}[h]
\centering
\includegraphics[width=0.42\textwidth]{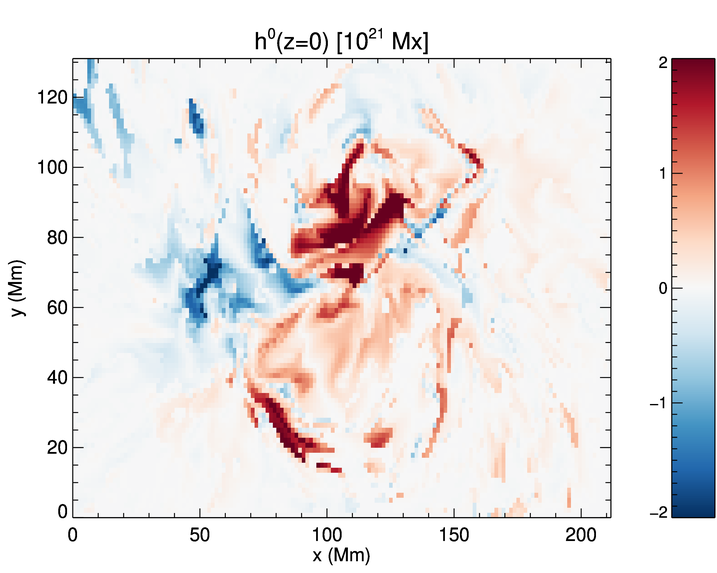}\\
\includegraphics[width=0.42\textwidth]{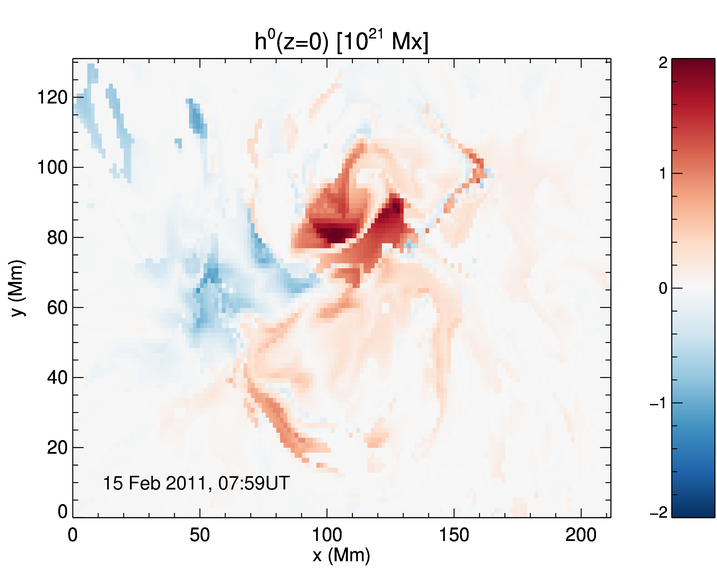}
\caption{Photospheric morphology of the RFLH in the unconstrained gauge (top), and in the Berger \& Field gauge (bottom), for AR 11158 on 07:59 UT of 15 February 2011. The two images exhibit a pixel-to-pixel linear correlation coefficient of 0.85.}
\label{gaugefig}
\end{figure}

We note here that the morphology of RFLH is not sensitive to the gauge used in its computation, at least in areas of high RFLH values. This can be seen in Fig.~\ref{gaugefig} where we compare the morphology obtained with the two gauges of Sect.~\ref{sect:method} for the snapshot at 07:59 UT of 15 February 2011. We see that the two distributions have similar morphologies, although they also exhibit some minor differences in small patches. The good agreement of the two maps is depicted in the high value of the pixel-to-pixel linear correlation coefficient between them, which is 0.85. The main difference between the two distributions is in the magnitude of RFLH which is by a factor of $\lesssim 2$ smaller in the Berger \& Field gauge. The morphology in this gauge is also more smooth because of the Coulomb gauge used in the computation of the vector potentials. Similar differences exhibit the other snapshots as well.

\subsection{Flare-related changes during the X2.2 flare}
\label{sect:res2}

\begin{figure}[h]
\centering
\includegraphics[width=0.45\textwidth]{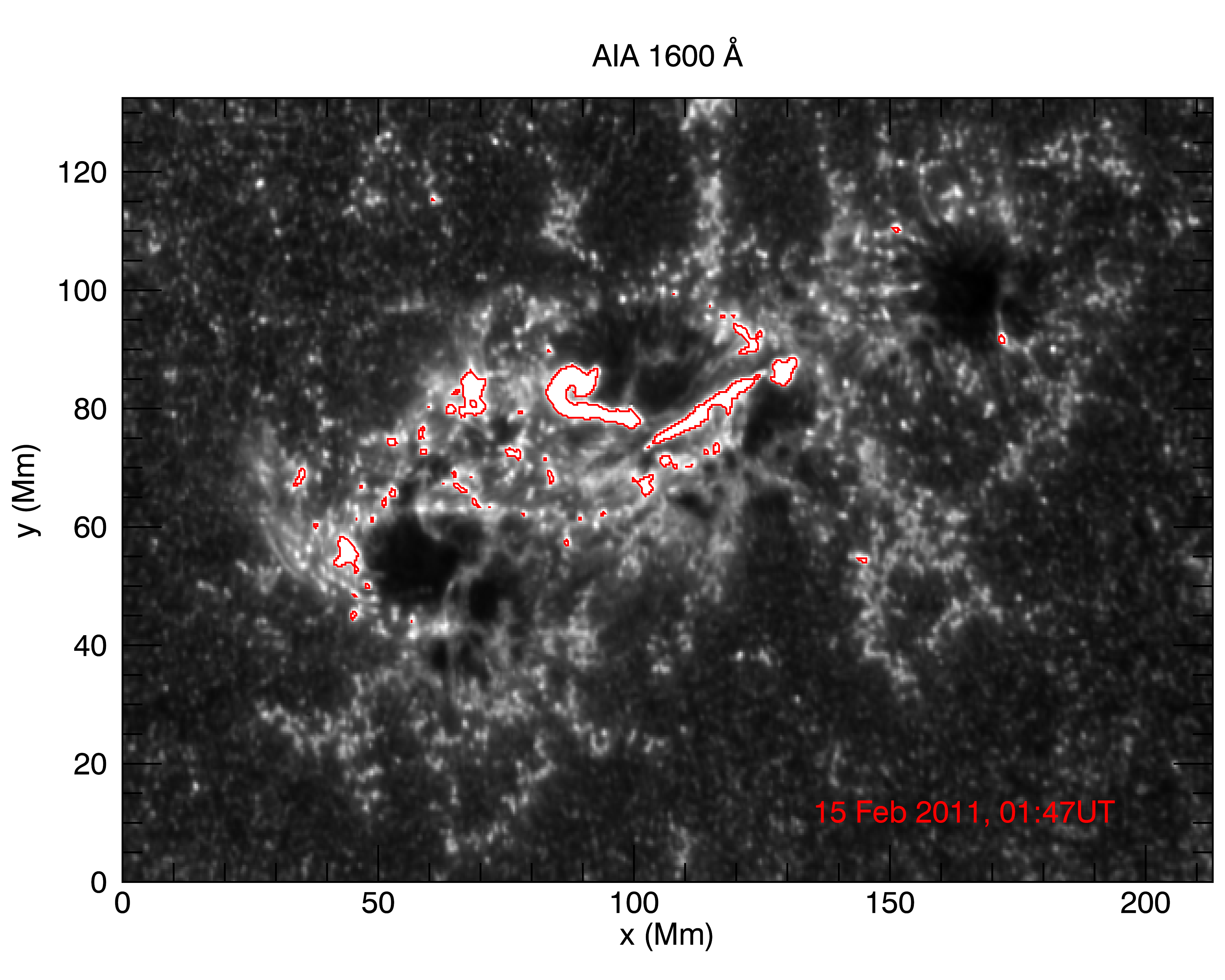}\\
\includegraphics[width=0.45\textwidth]{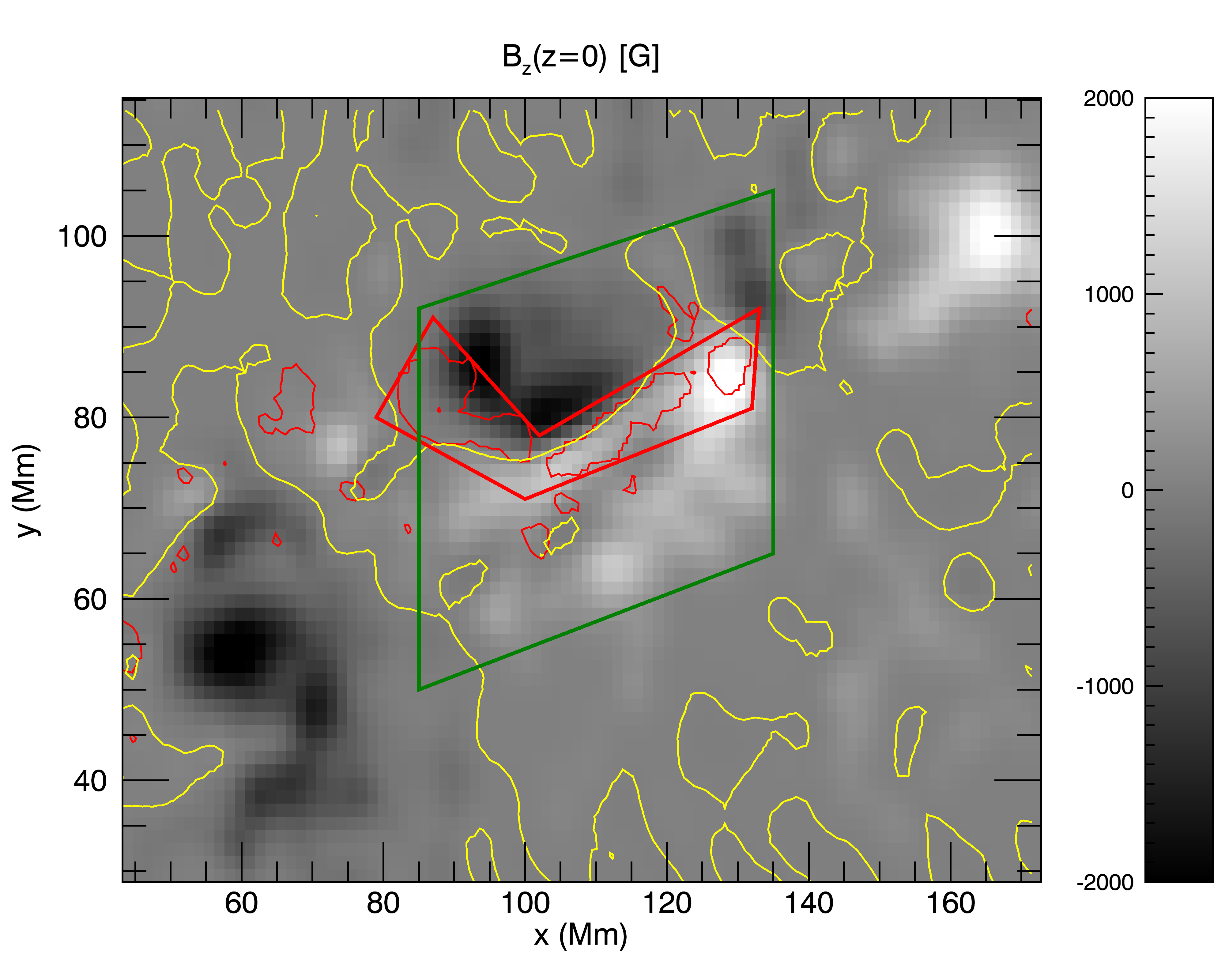}
\caption{AIA image at 1600~\r{A} of the same AR region shown in Fig.~\ref{flhfig1}, saturated at the intensity of 1180, and overplotted with the contours at the same value, at the beginning of the X2.2 flare of AR 11158 (top). Zoom of the vertical magnetic field 2D map at the same time, with the two regions of interest (bottom). Red contours are the same as in the top panel, only rebinned to the resolution of the magnetic field image. Yellow contours correspond to the magnetic field value $B_z=0\,{\rm G}$.}
\label{dataplot}
\end{figure}

We now focus on the X2.2 flare of AR 11158 of 15 February 2011. The flare's onset time was 01:44 UT and its peak time was 01:56 UT. Using observations from SDO's AIA instrument at 1600~\r{A} we are able to identify two regions related to the occurence of the flare. The first is the location of the flare ribbons which are shown in the top panel of Fig.~\ref{dataplot}, at the time of their maximum extent (01:48 UT) before the saturation of the image from the flare emission. According to the standard model for the 3D flares \citep{janvier13}, the flare ribbons correspond to the footprints of the quasi-separatrix layer (QSL) that separates the flux rope from the surrounding, confining field. This image resulted after correcting the original AIA image with standard methods (solar software's \texttt{aia$\_$prep} routine), and transforming it to the same projection and field of view (FOV) as the HMI images of Fig.~\ref{flhfig1}, following the recipes of \citet{sun13}. The latter procedure slightly deforms the shape of the image, but this is minimal due to the proximity of the AR to the solar disk centre. We approximate the region corresponding to the central, highest-intensity contour with the red polygon shown in the bottom panel of Fig.~\ref{dataplot}. The second region of interest includes the central polarities of the AR where the flare originated during its peak (as identified by eye inspection of AIA's images), although nearly the whole AR participated in this flare. This region is shown as a green box in the bottom panel of Fig.~\ref{dataplot}, and it encompasses the area covered by the ribbons when they reach their maximum extent.

\begin{figure*}[h]
\centering
\includegraphics[width=0.98\textwidth]{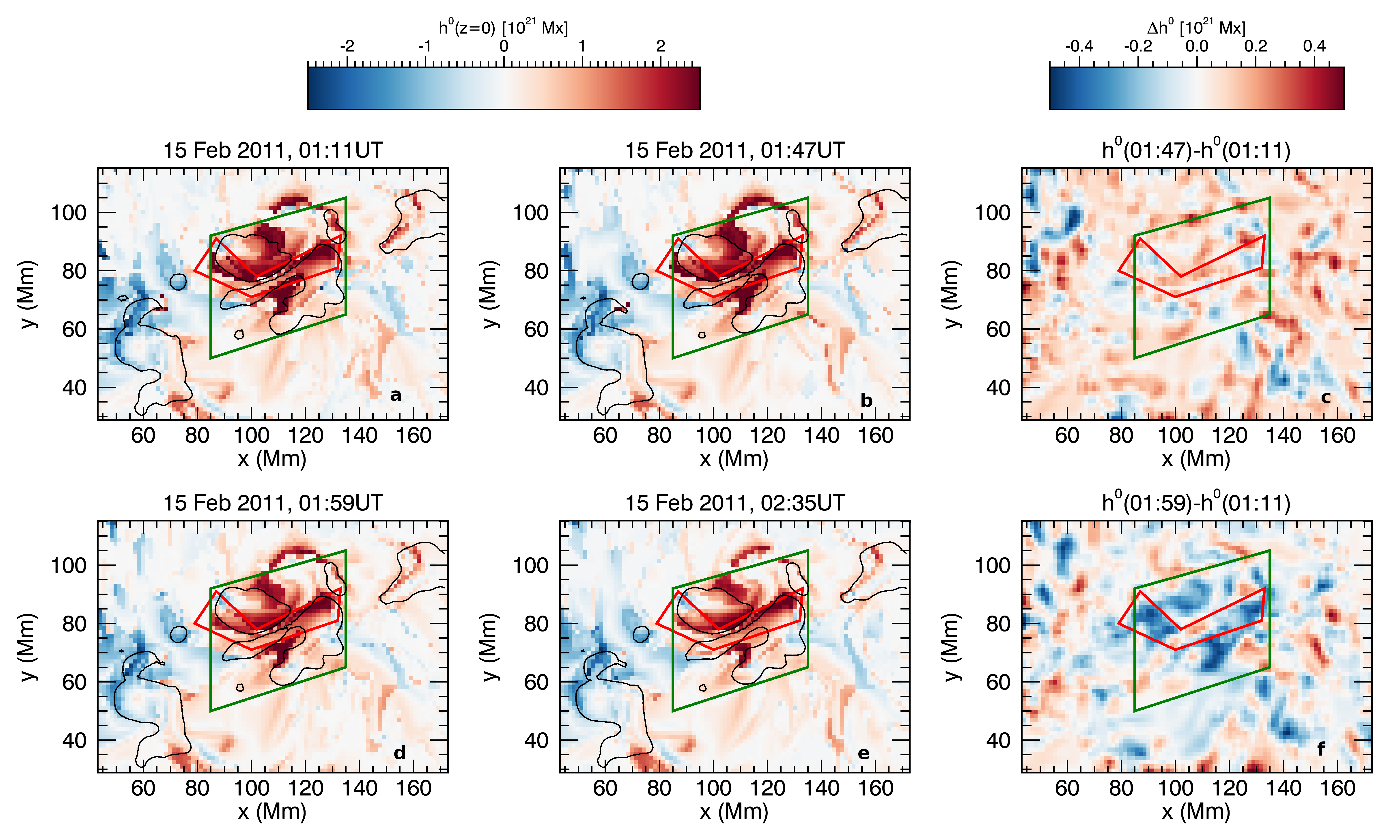}
\caption{Selected snapshots of the zoomed-in photospheric morhology of RFLH around the X2.2 flare of AR 11158 (panels a, b, d, e). Also shown are the two boxes defined in Fig.~\ref{dataplot}, and the contours of the vertical magnetic field values at $|B_z|=500\,{\rm G}$.\\
Difference images of the photospheric RFLH at the beginning of the flare (at 01:47 UT, panel c) and after the flare (at 01:59, panel f), with respect to the corresponding pre-flare image at 01:11 UT. The boxes are the same as in the rest panels.}
\label{dataplot2}
\end{figure*}

In the panels (a), (b), (d), and (e) of Fig.~\ref{dataplot2} we show the evolution of the RFLH distribution in the corresponding areas around the X-class flare. We note that the overall morphology has small variations in a two-hours interval around the flare. The most important difference is the large decrease in the RFLH magnitude between the snapshots at 01:47 UT and 01:59 UT, i.e., during the maximum of the flare. This is more evident in panels (c) and (f) where the difference of the images at 01:47 UT and at 01:59 UT with the image at 01:11 UT is shown, respectively. While the first difference image (panel c) shows small variations of both signs in the whole FOV, the second image (panel f) depicts a clear decrease in RFLH which is contained in the green box.

The use of RFLH thus enables us to identify the locations where helicity is lost from the AR during the flare and the respective eruption, and to also calculate its value. It is also interesting that the area of the RFLH decrease includes the location of the flux rope, and so we can deduce that a part of the flux rope was expelled during the flare. In other words, we are able to verify that the CME source coincides with the wider flux rope region. This conclusion agrees with many previous studies \citep[e.g.,][]{zhang12,patsourakos13,nindos15,chintzoglou15,nindos20}.

\begin{figure}[h]
\centering
\includegraphics[width=0.48\textwidth]{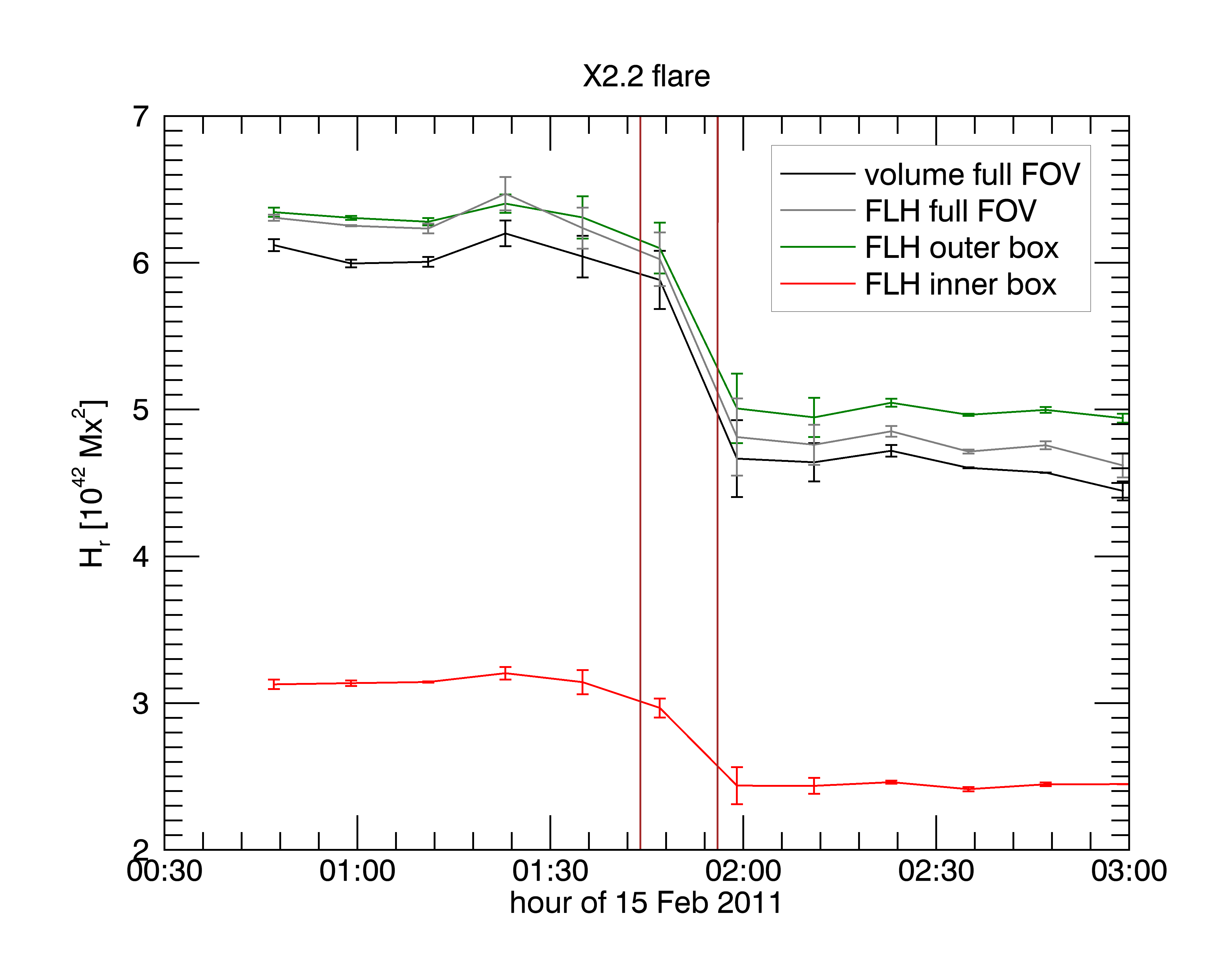}
\caption{Evolution of relative helicities in AR 11158 during a two-hours interval centered on the X2.2 flare. Shown are the relative helicities as calculated by the volume method (Eq.~(\ref{helr}), black line), and by the RFLH method using Eq.~(\ref{flhdef0}) in the whole AR (grey line), and in the two boxes of Fig.~\ref{dataplot} (red and green lines). The error bars are computed from the difference of the curves from their five-point moving averages. The vertical lines denote the onset and peak times of the X-class flare.}
\label{boxplot}
\end{figure}

In Fig.~\ref{boxplot} we show the evolution of various helicity budgets around the X2.2 flare. The black and gray curves are a zoom-in around the time of the X2.2. flare of the same helicity budgets as calculated by the volume and the RFLH methods for the full magnetogram, shown in Fig.~\ref{helplots}. We also show the errors stemming from the difference of the helicity curves from their five-point moving averages, which correspond to one-hour intervals. We use this type of error as an indication for the actual errors, since the computationally-deduced ones, by the method of \citet{moraitis14}, are very low.

In addition, Fig.~\ref{boxplot} depicts the helicity budgets calculated by the RFLH method for the two boxes of Fig.~\ref{dataplot}, as the green and red lines. In calculating these, we employ Eq.~(\ref{flhhel}) but consider only those field lines that have footpoints inside the respective areas.
		  
Figure \ref{boxplot} reiterates the good agreement of $\lesssim 5\%$ between the helicities obtained by the volume and the RFLH methods (black and grey curves respectively). Both curves exhibit a 25\% drop in helicity during the flare, which corresponds to $1.48\,10^{42}\,\mathrm{Mx}^2$ for the volume method, and to the slightly larger $1.56\,10^{42}\,\mathrm{Mx}^2$ for the RFLH one. The drop in the volume helicity is larger than the computed errors and so it is not an artifact but a real drop, a conslusion that holds for the other helicity curves as well. Similar sharp changes in other parameters related to the magnetic field have been reported previously for the same AR \citep{liu12n,sun17}.

Another conclusion that can be drawn from Fig.~\ref{boxplot} is	that the helicities obtained by the RFLH method for the full FOV and for the green box (grey and green curves respectively) agree to a large extent. This can be explained with the majority of the AR's helicity being contained in the green box, a conclusion also supported by the RFLH morphology in Fig.~\ref{dataplot2}. The agreement is more pronounced before the flare, when the two curves agree to within 0.5\%, while after the flare this falls to 5\%, which should be attributed to the rearrangement of the magnetic field during the flare, and to a lesser extent, to the evolution of the arc-shaped RFLH structure just above the green box in Fig.~\ref{dataplot2}. Finally, the drop in helicity coming from the green box is 20\%, a bit less than in the full FOV, and it corresponds to a value of $1.35\,10^{42}\,\mathrm{Mx}^2$.
		  
Focusing next to the red box that approximates the flare's ribbons we note that the helicity contained in it is the half of the entire AR. The relative drop in helicity during the flare is similar to the other curves, $\sim 22$~\% of its pre-flare value; its absolute value however is much smaller, corresponding to $7.1\,10^{41}\,\mathrm{Mx}^2$. This small relative drop indicates that the region stays non-potential even after the expulsion of the flux rope. On the other hand, the higher (absolute) helicity drop within the green box suggests that the effect of the CME is to change the entire surrounding of the flux rope. An alternative explanation could be that the eruption changes the mutual helicity between the systems of the flux rope and the surrounding field, in which case, the RFLH would reflect the non-local nature of helicity. This would also relate to the mutual-to-self conversion of helicity \citep{tgl13}, which was suggested for the same active region.

\subsection{Relation with the eruption of February 15}
\label{sect:res3}

The X2.2 flare of AR 11158 on February 15 was an eruptive one, that is, it was accompanied by a CME. This was observed with STEREO's COR1 coronagraph at 2:05 UT of February 15, and it had a peak velocity of $\sim 1300\, \mathrm{km}\,\mathrm{s}^{-1}$ \citep{maricic14}. The ejected structure took a part of the AR's helicity which, according to the previous section, should be positive (right-handed) and no more than the maximum estimated helicity decrease of $\sim 1.5\,10^{42}\,\mathrm{Mx}^2$. The CME was recorded three days later, starting at February 18 around 20:00 UT, as an Interplanetary Coronal Mass Ejection (ICME), using in situ data at the Lagrange point L1 from the \textit{Wind} spacecraft \citep{maricic14,lepping15}. This ICME was also listed in several other catalogs \citep{richardson10,chi16,hess17} at a similar time interval.

From the fitting of the ICME with a linear force-free magnetic cloud (MC) model, \citet[][Table 2 therein]{lepping15} find a value for the axial magnetic field strength of $B_0=11.4\,\mathrm{nT}$, and for the radius of the MC, $r_0=0.06\,\mathrm{AU}$. Further assuming the arbitrary, but consistent however with energetic electron recordings in ICMEs \citep{kahler11}, length for the MC of $L\sim 2\,\mathrm{AU}$, one can estimate the helicity of the ICME to $H_r\sim 2\,10^{41}\,\mathrm{Mx}^2$, following the equations in \citet{dasso06}. This value is significantly smaller than the entire AR's helicity, and also of the helicity contained in the two boxes. The comparison with the corresponding helicity drops shows that the MC's helicity is a factor of $\sim 7$ smaller than the helicity lost from the whole AR, or from the green box area. It is closer however to the helicity lost in the flare ribbons' area (red box), but still, a factor of $\sim 3$ smaller. A problem with the helicity reported in \citet{lepping15} is that this is left-handed, i.e., it has opposite sign compared to the AR. Moreover, as discussed by \citet{lepping15}, the quality of the MC fitting for this ICME is poor, exhibiting convergence issues. Therefore, this ICME does not represent the most optimal case for comparisons with solar observations. Despite all the uncertainties in the estimation of the MC's helicity, the absolute value obtained is consistent with the upper limit derived from the helicity drop in the whole AR.

\section{Discussion}
\label{sect:discussion}

This work dealt with the first detailed study of relative field line helicity in a solar active region. The target AR 11158 exhibited intense activity during the four-days study interval. The computation of RFLH was based on the relevant recent developments, and on a high-quality NLFF model for the AR's coronal magnetic field, thus ensuring reliable helicity estimations.

The examination of RFLH's photospheric morphology showed that this is quite different than the respective distributions of the magnetic field and of the electrical current. Furthermore, the morphology of RFLH was shown not to be sensitive on the employed gauge in its computation. The total AR helicity as estimated by the RFLH agrees to a large extent with the more accurate, volume helicity method, as was already demonstrated before.

Our results are consistent with the picture drawn by other works on AR 11158, e.g., that of \citet{jing12}. These authors find a similar helicity evolution, both overall, and during the X-class flare. The helicity of the AR exhibited a sharp decrease during the X2.2 flare, $\sim 25\%$ of its pre-flare value. A similar sharp drop was reported in \citet{jing12}, which, moreover, resulted from the respective drop in the right-handed helicity. This is exactly what the RFLH photospheric maps of Sect.~\ref{sect:res2} reveal. 

The main advantage of using RFLH is the additional spatial information that it can provide about the locations where helicity is more important, and its respective values. In the case of the X2.2 flare of AR 11158, we found that the decrease in helicity is coming mostly from the wider flux rope area. Additionally, by examining two observationally-deduced, flare-related regions, we were able to monitor the individual helicity evolution in each of them. The helicity drop during the flare in these regions did not reveal a clear relation with the helicity of the observed ICME, although there are many uncertainties in the derivation of the latter.

As a conclusion, this work pointed out the usefulness of RFLH in solar applications. The spatial information provided by RFLH can help in situations such as the one examined in this paper, i.e., in the determination of the pre-eruptive structures of coronal mass ejections. A possible next step in the study of RFLH would be to perform similar analysis to a number of ARs with different levels of activity, and different characteristics, and juxtapose them with the helicity of the associated ICMEs at 1~AU or even much closer in the corona or inner heliosphere with \textit{Parker Solar Probe} \citep{psp} and \textit{Solar Orbiter} \citep{solo20}. This would enable the examination of the behaviour of RFLH in different environments, and of the possibility of discriminating solar activity with the help of relative field line helicity.

\begin{acknowledgements}
The authors thank the referee for carefully reading the paper and providing constructive comments. This research is co-financed by Greece and the European Union (European Social Fund -- ESF) through the Operational Programme ``Human Resources Development, Education and Lifelong Learning'' in the context of the project ``Reinforcement of Postdoctoral Researchers -- 2$^{nd}$ Cycle'' (MIS-5033021), implemented by the State Scholarships Foundation (IKY). It also benefited from the discussions within the international team ``Magnetic Helicity in Astrophysical Plasmas'' which was supported by the International Space Science Institute. The authors also thank J. Thalmann for providing the NLFF reconstruction of the AR's magnetic field.
\end{acknowledgements}

\bibliographystyle{aa}
\bibliography{refs}

\end{document}